\documentclass[pra,onecolumn,11pt,amsmath, superscriptaddress,notitlepage,showpacs]{revtex4-1}
\usepackage{amsmath,amsfonts,amssymb,amstext,amscd,amsthm,dsfont,bbm,hyperref,natbib}
\usepackage{physics}
\usepackage{color}
\usepackage{soul,xcolor}
\usepackage{gensymb}
\hypersetup{
    colorlinks = true,
    citecolor=blue,
    linkcolor = red,
    anchorcolor = red,
    citecolor = blue,
    filecolor = red,
    pagecolor = red,
    urlcolor = red,
    }
\usepackage[normalem]{ulem}
\usepackage{graphicx}
\usepackage{bbold}
\usepackage{braket}
\usepackage{placeins}

\newtheorem{Theorem}{Theorem}
\newtheorem{Lemma}{Lemma}

\begin{document}
\title{Measure of positive and not completely positive single-qubit Pauli
    maps}
	\author{Vinayak Jagadish}
	\email{jagadishv@ukzn.ac.za}
	\affiliation{Quantum Research Group, School  of Chemistry and Physics,
		University of KwaZulu-Natal, Durban 4001, South Africa}\affiliation{ National
		Institute  for Theoretical  Physics  (NITheP), KwaZulu-Natal,  South
		Africa}
			\author{R. Srikanth}
	\email{srik@poornaprajna.org}
	\affiliation{Poornaprajna Institute of Scientific Research,
		Bangalore- 560 080, India}
	\author{Francesco Petruccione}
	\email{petruccione@ukzn.ac.za}
	\affiliation{Quantum Research Group, School  of Chemistry and Physics,
		University of KwaZulu-Natal, Durban 4001, South Africa}\affiliation{ National
		Institute  for Theoretical  Physics  (NITheP), KwaZulu-Natal,  South
		Africa}
	
\begin{abstract} 
The time evolution of an initially  uncorrelated  system  is governed  by  a
completely positive (CP)  map. More generally, the  system may contain
initial (quantum) correlations with an  environment, in which case the
system evolves according to a not-completely positive (NCP) map. It is
an interesting  question what  the relative measure  is for  these two
types of maps within the set of  positive maps.  After indicating the scope of the full problem of computing the true volume for generic maps acting on a qubit, we study the case of Pauli channels in an abstract space whose elements represent an equivalence class of maps that are identical up to a non-Pauli unitary. In this space, we show
	that the volume of NCP maps is twice that of CP maps. 
\end{abstract}
\maketitle

\section{Introduction}
The elementary units of any quantum information processing task are positive maps acting on the system of interest. The time-evolution of open quantum systems are not unitary as opposed to that for closed systems. The temporal evolution of open quantum systems~\cite{petruccione} are described by a dynamical map~\cite{sudarshan_stochastic_1961} acting on the set of states of the system of interest. The dynamical map should be linear, hermiticity and trace preserving and preserve the positivity for the states on which it acts on. A stronger version of positivity, named complete-positivity (CP) is often demanded for dynamical maps. Based on the phenomenological theory of relaxation of spins, it was argued that complete positivity was not a necessary requirement~\cite{simmons_completely_1981,raggio_remarks_1982,simmons_another_1982}. A clarification to the issue of complete positivity was brought forward in the series of papers by Pechukas and Alicki based on the concept of linear assignment maps~\cite{pechukas_reduced_1994,alicki_comment_1995,pechukas_pechukas_1995}. For initially entangled system and environment, it was shown
that the reduced dynamics can be not completely positive (NCP)~\cite{jordan_dynamics_2004,omkar2015operator} and the notion of positivity and compatibility domains were proposed~\cite{jordan_dynamics_2004}.

An interesting issue here would be the volume of different classes of qubit channels, which would quantitatively depend on specific parametrization schemes and corresponding measures.   Some related  work here: the volume of Pauli channels (or, generalized depolarizing channels) simulable with a one-qubit environment within the full space of Pauli channels is considered in~\cite{narang2007}. A similar calculation for amplitude damping channels is addressed in~\cite{jung2008}.  Ref.~\cite{lovas2018} employs the Lebesgue measure to compute the fraction of unital channels in the space of all qubit channels.  Ref.~\cite{Karo-2004}, using parametrization scheme developed in~\cite{karolbook} considers the volume of CP Pauli channels within the space of all CP unital qubit channels. Ref.~\cite{Szarek-2008}, employing the Choi-Jamiolkowski isomorphism between channels and extended states~\cite{Jamiolkowski-1972}, derives the Hilbert-Schmidt (Euclidean) volumes of the set of positive maps acting on a qudit, and its nested subsets of decomposable maps, CP maps and superpositive maps.

Here, we consider the question of the volume of CP maps within the space of positive maps acting on a qubit.  To the best of our knowledge, a volumetric analysis of any class of quantum channels inclusive of NCP maps has not been addressed os far.  We are interested in addressing the relative volume of positive and completely positive trace-preserving maps acting on a qubit. After first considering the general problem, we highlight specific technical issues in implementing the task. In particular, these challenges relate to finding a suitable parametrization for channels, and furthermore, characterizing the space of all NCP maps. We show that these constraints require us to consider the more limited case of Pauli CP maps within the space of Pauli positive maps.

The paper is organized as follows.  In Sec. \ref{truevol}, we discuss the preliminaries and address the issue with the calculation of a volume measure in the space of generic CP maps acting on a qubit. After pointing out the difficulty in handling this general situation, we  look into the case of unital maps in the associated Choi matrix representation in Sec. \ref{dynamicalunital}. The geometrical ideas and a result on the eigenvalue spectra of NCP maps acting on a qubit are given in Sec. \ref{geometry}. We address the reason as to why we further restrict our discussion to the case of Pauli channels. Based on the eigenvalues of the Choi matrix representing the positive maps, we propose a measure  of dynamical maps  acting on a qubit, motivate it, and prove our main results in Sec. \ref{measure}.   We conclude in Sec. \ref{conclusion}.
  \section{Choi Jamiolkowski Isomorphism and Volume of the Space of Qubit Channels}
  \label{truevol}
  Let  $\mathcal{E}$ be  a positive  map acting  on a
  qubit, represented by the state
\begin{equation}
  \label{eq:1q1}
  \rho = \frac{1}{2}(\mathbbm{1} + a_i \sigma_i) = \frac{1}{2} \left( \begin{array}{cc}1 + a_3 & a_1 - \imath a_2\\ a_1 + \imath a_2  & 1-a_3 \end{array} \right),
\end{equation}
where vector ${\bf  a} = (a_1 \,,\,  a_2 \,,\, a_3)$,
  with $|{\bf a}| \leq 1$, is the  Bloch vector. The space of all one
qubit states corresponds to the set of all the points on or inside the
``Bloch   ball''\index{Bloch   sphere}   which  is   the   unit   ball
in the space $\mathbb{R}^3$ parametrized by the axes
$a_1$, $a_2$ and $a_3$. 

The map $\mathcal{E}$ can be represented by a 4 dimensional Hermitian matrix, usually referred to as the Choi matrix~\cite{choi_completely_1975}. If the Choi matrix is positive, the map is CP and else NCP. The trace of the Choi matrix acting on a qubit is 2. Apart from the trace, the Choi matrix is equivalent to a valid state in 4 dimensions. This is the isomorphism between maps and states, usually referred to as the Choi-Jamiolkowski (CJ) isomorphism in literature. The required measure of one-qubit channels will then be the measure of two-qubit states. The two main technical issues we are confronted with is (1) to identify a proper parametrization for the volume of two-qubit states; (2) finding the proper generalization of the parameters to include NCP maps, which correspond to non-positive two-qubit states.

The density matrix in 4 dimensions can be expressed in terms of the generators of SU(4) Lie algebra. Following~\cite{tilma2002}, any $4\times 4$ density matrix, $\tilde{\rho}$ can be parametrized using the generators of SU(4) Lie algebra, $\{\Lambda_i\}$ with 12 Euler angles $\alpha_i$ and three rotation angles $\theta_i$ as follows
\begin{widetext}
\begin{eqnarray}
\label{uberrho}
\tilde{\rho }&=& e^{i\Lambda_3 \alpha_1}e^{i\Lambda_2 \alpha_2}e^{i\Lambda_3 \alpha_3}e^{i\Lambda_5 \alpha_4}e^{i\Lambda_3 \alpha_5}e^{i\Lambda_{10} \alpha_6}e^{i\Lambda_3 \alpha_7}e^{i\Lambda_2 \alpha_8}
e^{i\Lambda_3 \alpha_{9}}e^{i\Lambda_5 \alpha_{10}}e^{i\Lambda_3 \alpha_{11}}e^{i\Lambda_2 \alpha_{12}} \nonumber \\
& &\times (\frac{1}{4}\mathbb{1}_4+\frac{1}{2}(-1+2a^2)b^2c^2*\Lambda_3 +
\frac{1}{2\sqrt{3}}(-2+3b^2)c^2*\Lambda_8+\frac{1}{2\sqrt{6}}(-3+4c^2)*\Lambda_{15})\nonumber\\
& &\times e^{-i\Lambda_{2} \alpha_{12}}e^{-i\Lambda_{3} \alpha_{11}}e^{-i\Lambda_{5} \alpha_{10}}e^{-i\Lambda_{3} \alpha_{9}}
e^{-i\Lambda_{2} \alpha_{8}}e^{-i\Lambda_{3} \alpha_{7}}e^{-i\Lambda_{10} \alpha_{6}}e^{-i\Lambda_{3} \alpha_{5}}e^{-i\Lambda_{5}
\alpha_{4}}e^{-i\Lambda_{3} \alpha_{3}}e^{-i\Lambda_{2} \alpha_{2}}\times e^{-i\Lambda_{3} \alpha_{1}}
\end{eqnarray}
\end{widetext}
where
\begin{eqnarray}
\label{factorlist}
a^2 &=& \sin^2(\theta_1),\nonumber\\
b^2 &=& \sin^2(\theta_2),\nonumber\\
c^2 &=& \sin^2(\theta_3),
\end{eqnarray}
with the ranges for the 12 $\alpha$ parameters and the 
three $\theta$ parameters given by
\begin{eqnarray}
0 \le \alpha_1,\alpha_3,\alpha_5,\alpha_7,\alpha_9,\alpha_{11} \le
\pi, \nonumber \\
0 \le \alpha_2,\alpha_4,\alpha_6,\alpha_8,\alpha_{10},\alpha_{12} \le
\frac{\pi}{2}, \nonumber 
\end{eqnarray}
\begin{eqnarray}
\frac{\pi}{4} \le \theta_1 \le \frac{\pi}{2}, \quad &\nonumber\\
\cos^{-1}(\frac{1}{\sqrt{3}}) \le \theta_2 \le \frac{\pi}{2}, \quad &\nonumber\\
\frac{\pi}{3} \le \theta_3 \le \frac{\pi}{2}.
\end{eqnarray}
Based on this parametrization, a volumetric analysis could be performed based on the Haar measure of the SU(4) states. In other words, via the CJ isomorphism the volume of the space of CP/NCP maps can be evaluated in terms of positive/negative Choi matrices, in principle.  But, as we can see, a calculation of volume in terms of the above mentioned Euler-angle parametrization is tedious and an analytical expression is hard to be obtained, let alone for NCP maps.

Therefore, we ask the question whether a volume issue could be addressed for a sub-class of  maps acting on a qubit. A rather limited, and yet very important and large set of maps is the unital maps (which map a unit matrix back to itself).

\section{Unital Maps Acting on a Qubit}
\label{dynamicalunital}
The general form of the Choi matrix $\tilde{B}$ acting on a qubit which is unital and trace-preserving can be parametrized as follows. This is obtained by considering a 4 dimensional Hermitian matrix such that $\mathrm{tr}_{1}\tilde{B} = \mathbbm{1} = \mathrm{tr}_{2}\tilde{B}$. Note that $a$ is real and $x,y,z$ and $w$ are in general complex.
\begin{equation}
\label{equnitalgeneral}
\tilde{B}= \left(
\begin{array}{cccc}
 a & x & y & z \\
 x* & 1-a & w & -y \\
 y* & w* & 1-a & -x \\
 z* & -y* & -x* & a \\
\end{array}
\right).
\end{equation}
Unfortunately, even in this case, although computing the volume of maps is comparatively easier than in the general case, there seems to be no way yet to bound the full volume of maps including the NCP case. We shall thus require a further restriction of the  sub-class of Pauli channels, discussed below, for which we can use an existing result to compute the volume of both CP and NCP maps in an abstract, albeit well-motivated, space. The reason for not using the general form as in Eq. (\ref{equnitalgeneral}) is explained later in the manuscript.

Let us consider the following form of the Choi matrix $\mathcal{B}$ representing a trace-preserving $\mathcal{E}$ on a qubit,
\begin{equation}
\label{eqmap}
\mathcal{B} = \frac{1}{2} \left(
\begin{array}{cccc}
1+ t_3+x_3 & t_1-\imath t_2 & 0 & x_1+x_2 \\
 t_1+\imath t_2 & 1-t_3-x_3 & x_1-x_2 & 0 \\
 0 & x_1-x_2 & 1+t_3-x_3 & t_1-\imath t_2 \\
 x_1+x_2 & 0 & t_1+\imath t_2 & 1-t_3+x_3 \\
\end{array}
\right).
\end{equation}
This is obtained  by considering the action of  the map $\mathcal{E}$
on the basis $\{\mathbbm{1}, \sigma_{i = 1, 2, 3}\}$ as
\begin{eqnarray}
\label{eqmapdef}
\mathcal{E}(\mathbbm{1}) &=& \mathbbm{1} + \sum_{i = 1}^{3} t_{i} \sigma_{i},\nonumber\\
\mathcal{E}(\sigma_{i}) &=& x_{i} \sigma_{i},
\end{eqnarray}where  $\mathbbm{1}$ is the identity matrix in two dimensions and $\sigma_i$ are the familiar Pauli matrices:
\[ \sigma_1 = \left( \begin{array}{cc}0 & 1 \\ 1 & 0 \end{array} \right) \quad ; \quad   \sigma_2 = \left( \begin{array}{cc}0 & -\imath \\ \imath & 0 \end{array} \right) \quad ; \quad \sigma_3 = \left( \begin{array}{cc}1 & 0 \\ 0  & -1 \end{array} \right). \]
This means that the map scales each of the three independent directions with scaling factors $x_i$ and translates $t_i$ along the three directions. The dynamical matrix Eq. (\ref{eqmap}) can be easily written down by inspection, by using the definition of the action of the map as in Eq. (\ref{eqmapdef}) on Eq. (\ref {eq:1q1}). Eq. (\ref{eqmap}) represents a dynamical map acting on a qubit with affine shifts. Such a map is refereed to as non-unital. Note that  $\displaystyle \mathrm{tr}_{2}\mathcal{B}  = \mathbbm{1}$, which means that the map is trace-preserving.

Restricting ourselves to the  case of Pauli channels, we
  set $t_j \equiv 0$ in Eq. (\ref{eqmap}), so that
\begin{equation}
\label{equnital}
  B= \frac{1}{2}\left( \begin{array}{cccc}
1+x_3 & 0 & 0 & x_1+x_2 \\ 0 & 1-x_3 & x_1 -x_2 & 0 \\ 0 & x_1 - x_2 & 1-x_3 & 0 \\ x_1+x_2 & 0 & 0 & 1+x_3
 \end{array} \right).
\end{equation}
The eigenvalues of $B$ are
\begin{eqnarray}
  \label{eq:1q6}
  \lambda_1 & = & \frac{1}{2} ( 1 + x_1 - x_2 - x_3), \nonumber \\
   \lambda_2 & = & \frac{1}{2} ( 1 - x_1 + x_2 - x_3), \nonumber \\
    \lambda_3 & = & \frac{1}{2} ( 1 - x_1 - x_2 + x_3), \nonumber \\
     \lambda_4 & = & \frac{1}{2} ( 1 + x_1 + x_2 + x_3).
     \end{eqnarray}

\section{Pauli Unital Maps, Positivity and Geometry}
\label{geometry}
 The set of maps parametrized as in Eq. (\ref{equnital}) is usually referred to as the Pauli channels. A Pauli channel has a Kraus representation as the convex combination of application of the four Pauli operators. However, the considerations of this and the next section would equally apply to any other sub-class of unital maps where the operators are a set of four orthogonal unitary operator basis elements.
With the assumption of unitality, the  map can be represented as an action on the Bloch vector as ${\bf
  a}^\prime = T {\bf  a}$ where $T$ is a $3\times 3  $ matrix. $T$ can
be diagonalized  and the  eigenvalues of $T$  are invariant  under the
unitary transformation. The matrix of eigenvalues is notated as $D$ in the subsequent discussions. For the unital map $B$ in Eq. (\ref{equnital}), the diagonal matrix, $D$  has entries $\vec{x} =  \{x_{1}, x_{2},  x_{3}\}$ which is equivalent to the fact that the $T$ matrix itself is diagonal.

The elements  of $D$  specify a
tetrahedron  $\mathcal{T}$ in  the  parameter  space $\{x_{1},  x_{2},
x_{3}\}$.  $\mathcal{T}$ is a simplex, given by the convex hull of the
points representing $\mathbbm{1}$ and  the Pauli matrices.  The points
forming the vertices of $\mathcal{T}$  correspond to unitary maps, the
edges and  faces represent two  and three operator  maps respectively.
Points   inside  $\mathcal{T}$   need  all   four  operators.    Thus,
$\mathcal{T}$   represents  the   set  of   all  Pauli   channels.  If
$\mathcal{E}$ has to  be completely positive, then  all $\Lambda_i$ in
Eq. (\ref{eq:1q6})  must be  positive semi-definite.  This  means that
the scaling parameters $x_i$ have to be such that
\begin{eqnarray} 
|1\pm x_3| \geq |x_{1}\pm x_{2}| &\Leftrightarrow \textrm{CP~map}, \nonumber \\
|1\pm x_3| \leq |x_{1}\pm x_{2}| &\Leftrightarrow \textrm{NCP~map}.
\label{eq:cond}
\end{eqnarray}

Applying the general map on a general density matrix for a qubit, i.e,
applying the map Eq. (\ref{eqmap}) in Eq. (\ref{eq:1q1}), one can see that
for positivity, $|x_i| \leq 1$.   Therefore, the space of positive and
trace preserving (PTP) maps form a unit cube, irrespective of whether the map is unital or not. This is physically meaningful, as positivity is a statement about the action of the map, unlike complete positivity which is a statement about the map itself. If the eigenvalues of $\mathfrak{B}$ are positive, then the map is complete positive (CP). If there is one eigenvalue which is negative, the map is not completely positive (NCP). We now state the relevant details and prove our results subsequently.

\begin{Theorem} 
\label{wolftheorem}(Wolf and Cirac~\cite{wolf_dividing_2008}),
For any positive map which is trace-preserving, the determinant of $T$
is contained in $[-1,1]$.
\end{Theorem}
Based on Theorem \ref{wolftheorem}, we state our result on the bound on the eigenvalues of the Choi matrix associated with a Pauli channel.
\begin{Lemma}
\label{lemma1}
The eigenvalues of  the Choi matrix, $B$ corresponding to  any positive Pauli channel
on a qubit which is trace preserving are bounded and cannot be greater
than 2 in absolute value.
\end{Lemma}
{\bf   Proof:} From Theorem \ref{wolftheorem},
we have that $\prod _{i} x_i \in [-1,1]$, from which it follows that
\begin{eqnarray}
 \forall_{i} x_i \in [-1,1].
\end{eqnarray}  
Consider  $\lambda_1$: in  view of  Eq. (\ref{eq:1q6}),  it assumes  the
largest absolute value  by setting $x_1 := \pm1$, $x_2  {:=} \mp1$ and
$x_3 {:=} \mp1$, with this value being 2. Repeating this for the other
eigenvalues $\lambda_i$, we find that:
\begin{equation}
\forall_i |\lambda_i | \le 2,
\end{equation}
meaning that any eigenvalue is bounded.

If the map is CP, given that the eigenvalues must sum to 2, it is very
clear  that  no eigenvalue  can  be  greater  than  2, since  all  are
positive.  But,  for NCP maps  (wherein the dynamical matrix  can take
negative eigenvalues), this result is non-trivial.
\hfill $\blacksquare$
\bigskip

For the form of  $\tilde{B}$ as in Eq. (\ref{equnitalgeneral}), the map in the affine form gives the $T$ matrix as
\begin{equation}
\label{tunitalnondiag}
T = \left(
\begin{array}{ccc}
 \mathrm{Re}[w+z] & -\mathrm{Im}[w+z] & 2  \mathrm{Re}[x] \\
  \mathrm{Im}[z-w] &  \mathrm{Re}[z-w] & 2  \mathrm{Im}[x] \\
 2  \mathrm{Re}[y] & -2  \mathrm{Im}[y] & 2 a-1 \\
\end{array}
\right).
\end{equation}
From the determinant of the $T$ matrix as in Eq. (\ref{tunitalnondiag}) and the eigenvalues of $\tilde{B}$ of Eq. (\ref{equnitalgeneral}), one can easily note that for certain choices of the parameters of $\tilde{B}$, the eigenvalues can blow up to very high values. It should be noted that the problem arises only for NCP maps, where the eigenvalues can take negative values. For CPTP maps, the eigenvalues are still bounded. This makes the characterization of the set of positive maps corresponding to general qubit unital maps difficult and hence we consider only the special class of Pauli unital channels as parametrized by Eq. (\ref{equnital}).
\section{Measure of the map}
\label{measure}

We now parametrize  all Pauli unital qubit maps in terms  of the eigenvalues
of  their  dynamical matrix,  i.e.,  the  set  of three  real  numbers
corresponding to the scaling factors $(x_1, x_2, x_3)$.  
  We define  the measure $\mu$ of  a set of maps as  the volume of
  the set in  the above parameter space  $\mathfrak{C}$ of eigenvalues
  of the dynamical  matrix.  That is, given set  $S \in \mathfrak{C}$,
  the measure  of $S$ is its  volume in Cartesian coordinates  up to a
  normalization factor $\kappa$
\begin{equation}
\mu(S) = \kappa \int\int\int_S dxdydz,
\label{eq:volume}
\end{equation}
where  $\kappa$  is fixed  so  that  $\mu(\mathfrak{C})=1$, i.e.,  the
measure of the  space of all positive trace-preserving  (PTP)  maps is unity. Later,  we shall find
that  $\kappa  = \frac{1}{8}$.  Note  that  this  space is  not  the
physical space.  Operationally,  the measure $\mu(S)$ of a  set $S$ of
maps in  this space  is the  probability with which  we would  pick an
element  of $S$  in the  set  of all  PTP
unital maps of a qubit.

Some
properties of this parameter space are noted below.
\begin{Lemma} 
In the space $\mathfrak{C}$ of all  Pauli PTP unital qubit maps, the measure
$\mu$ of unitaries is 0.
\end{Lemma}
{\bf Proof:} The measure of the tetrahedron 
$\mathcal{T}$ is, by definition, $\frac{8}{3}$. 
As noted earlier, the set $\mathcal{U}$ of unitaries correspond to points
forming the vertices of $\mathcal{T}$. As their Cartesian volume
is 0, it follows that $\mu(\mathcal{U})=0$. 
  \hfill $\blacksquare$
\bigskip

 Note that as the edges and faces correspond to maps with
  two  and   three  Kraus  operators,  even  these   have  zero
  measure. Only the set of  points inside $\mathcal{T}$, which require
  all four Kraus operators, have finite measure $\mu$.
\begin{Lemma}
 Each point  in the space $\mathfrak{C}$  corresponds to a map  up to a non-Pauli
unitary.
\end{Lemma}
{\bf Proof:}  Each point in  $\mathfrak{C}$ represents a
  map  parametrized  in  terms  of  the  eigenvalue  spectrum  of  its
  dynamical matrix.  Given  a density matrix $\rho$,  let its spectral
  decomposition be  $\sum_j \Lambda_j  \ket{j}\bra{j}$ and let  $U$ be
  any unitary  acting on it. Then,  $U \rho U^\dag =  \sum_j \Lambda_j
  \ket{(U)j}\bra{(U)j}$, where $\ket{(U)j}  \equiv U\ket{j}$. 
  
  Clearly,
  the eigenvalue  spectrum doesn't  change. It  thus follows  that the
  eigenvalues of  the dynamical matrix  are invariant when  rotated by
  any   given  unitary   acting  on   the  qubit.  This unitary equivalence can be used to establish an equivalence class of noise.   However, note that when the unitary is a Pauli operator, it has the effect of permuting the eigenvalues. Excluding the Pauli from the unitiaries that define our equivalence would entail that two channels are equivalent iff they differ by a unitary other than a Pauli operation. Thus, the set of unitary channels is represented by four vertices, corresponding to the Pauli qubit operations. \hfill
$\blacksquare$
\bigskip

In the scope of this  definition of parameter space $\mathfrak{C}$, we show that the ratios of the  volume of the space of CPTP and NCPTP
unital maps on a qubit represent the relative volume (measure).

\begin{Lemma}
The volume of Pauli unital NCPTP maps is twice as large as that of CPTP maps.
\end{Lemma} 
{\bf Proof:} Let $V_{PTP}, V_{CPTP}  ,V_{NCPTP}$ denote the volumes of
positive,  completely-positive   and  non-completely   positive  trace
preserving maps respectively in the above parameter space.  Obviously,
$V_{PTP} = V_{CPTP} + V_{NCPTP}$.  The  volume of space of PTP maps is
that  of the  cube with  each side  ranging in  $[-1,1]$ and is  8
units. By normalization, the measure of this cube is set
  to  1,  so  that  $\kappa$   in  Eq.  (\ref{eq:volume})  is  set  to
  $\frac{1}{8}$.   The  CPTP   maps,  determined   by  the   condition
  (\ref{eq:cond}),  are contained  in  the tetrahedron  $\mathcal{T}$,
  which is of volume $\kappa \frac{8}{3}=\frac{1}{3}$.  The difference
  in  volume  $\kappa(8  -   V_{CPTP})  =  16\kappa/3  =  \frac{2}{3}$
  corresponds to that  of the complementary space of  NCPTP maps. It
thus follows that the  volume of NCPTP maps is twice  as large as that
of CPTP maps.  \hfill $\blacksquare$
\bigskip

In addition to giving a  compact comparison of CP and NCP
  maps, this  result can be useful  for simulating PTP, CPTP  or NCPTP
  maps, in light of the bounds given in Eq. (\ref{eq:cond}). That is, a program
  may randomly  pick three reals  $x_j$ subject to  these constraints,
  and this would define a PTP map up to a unitary.
\section{Conclusions}
\label{conclusion}

For long,  the study  of open  systems remained  confined to  CP maps.
With  the  recognition  of  the importance  of  initial  correlations,
non-markovian dynamics~\cite{breuer_colloquium:_2016,li_concepts_2017}
and improved  experimental techniques, NCP  maps are now  studied ever
more actively~\cite{rivas_entanglement_2010}. This
  motivates the  question of how likely  an arbitrary trace-preserving
  dynamical map is likely to be unitary, CP or NCP. 
Exploiting the CJ isomorphism, we transform this to the question of comparing the volume of two-qubit positive and non-positive states. This turns out to be rather hard to handle for two reasons: (1) dealing with the 15-parameter space required to characterize this set; (2) fixing the exact limits when NCP maps are included. The first problem becomes considerably simplified when we restrict to the class of unital channels, but the second problem persists. We make a further restriction, to the class of Pauli unital channels. The resulting volumetric analysis should be applicable to any situation where the Pauli set is replaced by another unitary qubit operator basis. 
In the  context of qubit  dynamics we  defined a measure  of dynamical
maps based on the eigenvalues  of the dynamical matrix.  Physically, this corresponds to an abstract space where each element represents an equivalence of class of noisy channels up to a non-Pauli unitary.
In this framework, we have  investigated the relative measure of CP
and NCP Pauli unital  maps and showed that  the volume of NCP  maps is twice
that of CP maps. Future work may consider issues related to bounding the space of NCP maps. Another direction would be to explore the possibility for two-qubit parametrization other than Eq. (\ref{uberrho}), possibly one based on Bloch ball representation.

\section{Acknowledgements}
The work  of V.J.  and F.P.  is based upon  research supported  by the
South African Research  Chair Initiative of the  Department of Science
and  Technology and  National Research  Foundation.  R.S.   thanks the
Defense Research  and Development  Organization (DRDO), India  for the
support      provided       through      the       project      number
ERIP/ER/991015511/M/01/1692.
\bibliography{Affine}

\end{document}